\documentclass[11pt,twoside]{article}
\usepackage{asp2010}
\resetcounters

\begin{document}

\title{OB Stars in Stochastic Regimes}
\author{M. S. Oey$^1$, J. B. Lamb$^1$, J. K. Werk$^1$, and C. J. Clarke$^2$}
\affil{$^1$University of Michigan, Department of Astronomy, Ann Arbor, MI\ \ \ 48109-1042, USA}
\affil{$^2$Institute of Astronomy, Madingley Road, Cambridge CB3 0HA, UK}

\begin{abstract}
The highest-mass stars have the lowest frequency in the stellar IMF, and they are
also the most easily observed stars.  Thus, the counting statistics for OB stars
provide important tests for the fundamental nature and quantitative parameters
of the IMF.  We first examine some local statistics for the stellar upper-mass limit itself.
Then, we examine the parameter space and statistics for extremely sparse clusters
that contain OB stars, in the SMC.  We find that thus far, these
locally observed counting statistics are consistent with a constant
stellar upper-mass limit.  The sparse OB star clusters easily fall
within the parameter space of Monte Carlo simulations of cluster
populations.  If the observed objects are representative of their
cluster birth masses, their existence implies that the maximum stellar mass is
largely independent of the parent cluster mass.
\end{abstract}

\section{Introduction}

Since massive stars are exceedingly rare, any local, empirical determination
of their statistical properties is necessarily a stochastic problem.
This applies to determining the slope of the initial mass function
(IMF) and stellar upper-mass limit.  While the properties of massive
stars are inferred from models for both stellar atmospheres and
stellar evolution, we can begin to set constraints on the statistical
properties of the population once we believe we can make consistent,
systematic estimates of the stellar properties.  However, it is
important to bear in mind the systematic uncertainties inherent in
these determinations (Massey, these Proceedings).  Here, I will examine a few
stochastic analyses of local observations of the OB star population.  

\section{UP:  The Stellar Upper-Mass Limit}

If the IMF truly behaves as a probability density function, which is
indeed the way that we ordinarily assume that it does, then in
principle we can collect data from many different clusters and combine
them to increase the size of our sample to beat down stochastic noise.
We applied this technique to evaluate the stellar upper-mass limit
\citep{OeyClarke2005}, using data from local OB associations.
\citet{Massey1995} obtained uniform, high-quality spectroscopic
classifications of the upper IMF in a number of OB associations in the
Milky Way and Magellanic Clouds.  To evaluate the upper-mass limit, we
are interested only in clusters having ages $\lesssim 3$ Myr, which
are young enough so that we are observing these systems before the most
massive stars have expired as supernovae (SNe).  Eight objects studied by
Massey et al. (1995) qualify:  IC 1805, Berkeley 86, NGC 7380, NGC
1893, NGC 2244, Tr 14/16, LH 10, and LH 117/118.  These clusters
contain, respectively, 24, 10, 11, 19, 12, 82, 65, and 40 stars having
masses $\geq 10$ M$_\odot$, for a total of 263 stars. 
In addition to this sample, Massey \& Hunter (1998) used the same
techniques to obtain the census of the upper IMF in the R136a super
star cluster in the 30 Doradus complex in the LMC, using {\it
  HST}/STIS spectroscopy.  Their strong lower limit for the number of
stars having masses $\geq 10$ M$_\odot$ is 650.  Combining the samples of
ordinary OB associations and R136a yields a grand total of 913 stars.

\begin{figure}
\plotone{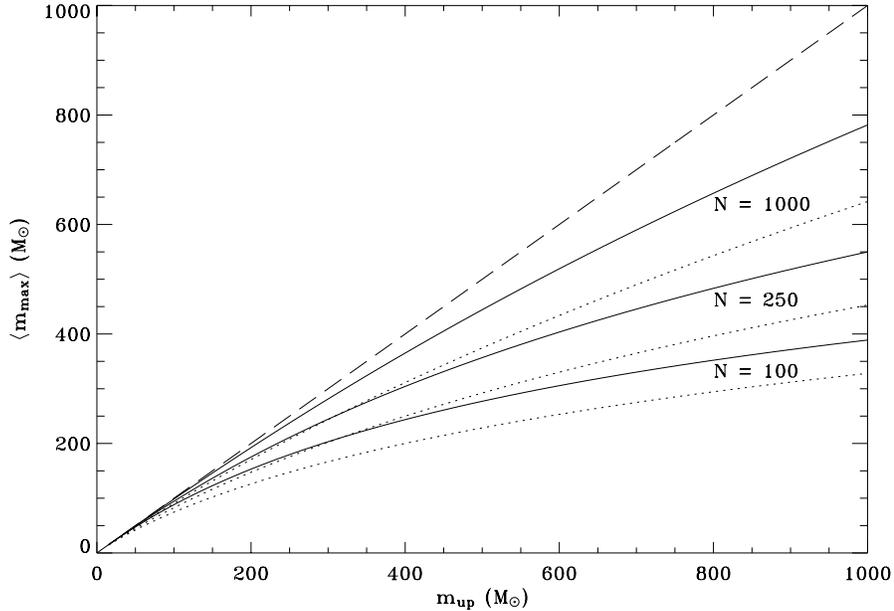}
\caption{Expected maximum mass $\langle m_{\rm max}\rangle$ vs 
  upper-mass limit $m_{\rm up}$ of the parent distribution,
for ensembles of $N_* = 100,$ 250, and 1000 stars having masses $\geq
10\rm M_\odot$.  Dotted lines assume the parent IMF has the form of a
truncated Salpeter power law, and solid lines assume the form in
equation~\ref{eqaltIMF}.  The dashed line shows the locus for $\langle
m_{\rm max}\rangle = m_{\rm up}$.
\label{werkfig}
}
\end{figure}

The maximum stellar mass in the entire sample is around 120 -- 150 M$_\odot$.  We
can compare this with the average expected maximum mass for an IMF
with a \citet{Salpeter1955} power-law slope $\gamma= 2.35$, for an assumed
upper-mass limit $m_{\rm up}$ of the parent distribution, and a given number
of stars $N_*$ in the ensemble.  As before, $N_*$ includes only stars
of mass $m\geq10$ M$_\odot$.  The expected maximum mass $\langle
m_{\rm max}\rangle$ is
given by:
\begin{equation}\label{eqmmax}
\langle m_{\rm max}\rangle = m_{\rm up} - 
   \int_{0}^{m_{\rm up}} \Biggl[\int_{0}^M \phi(m)\ dm\Biggr]^N \ dM   \quad .
\end{equation}
Figure~\ref{werkfig} shows the average expected $\langle m_{\rm
  max}\rangle$ as a function of $m_{\rm up}$ for three different
values of $N_*$ (dotted lines).
For a true upper-mass limit $m_{\rm up} = 1000$ M$_\odot$, an
ensemble having $N_*=250$ has an expected $\langle m_{\rm max}\rangle \sim 450$ M$_\odot$;
an ensemble having $N_*=1000$ has $\langle m_{\rm max}\rangle \sim 650$ M$_\odot$.  These
values of $N_*$ are similar to those for our observed combined samples
of Milky way and LMC associations, excluding and including R136a.  
We note that the values of $\langle m_{\rm max}\rangle$ do have some
dependence on the assumed form of the parent IMF.  The dotted lines in
Figure~\ref{werkfig} show the results from equation~\ref{eqmmax} for a
simple Salpeter power-law truncated at $m_{\rm up}$.  However, if we
adopt a ``softer'' truncation of the form,
\begin{equation}\label{eqaltIMF}
n(m)\ dm \propto \Biggl[\Biggl(\frac{m}{m_{\rm
      up}}\Biggr)^{-\gamma}-1\Biggr]\ dm \quad ,
\end{equation}
then instead equation~\ref{eqmmax} yields the solid lines in
Figure~\ref{werkfig}. 

In any case, as far as we know, no stars approaching these expected
$m_{\rm max}$ values on
the order of a few hundred M$_\odot$ are known to be observed, and the
example calculations assume a true $m_{\rm up}$ of merely 1000
M$_\odot$, much less infinity.  Indeed, inverting the argument, the
observed values of $m_{\rm max}\sim 150$ M$_\odot$ imply that the
parent $m_{\rm up}$ has a 
similar value.  \citet{Elmegreen2000} used this reasoning to estimate that
if $m_{\rm up}=\infty$, then somewhere in the Milky Way, there should be a
star having $m_{\rm max}$ = 10,000 M$_\odot$.  No star remotely approaching
this mass is suggested to have been seen.

\begin{figure}
\plotfiddle{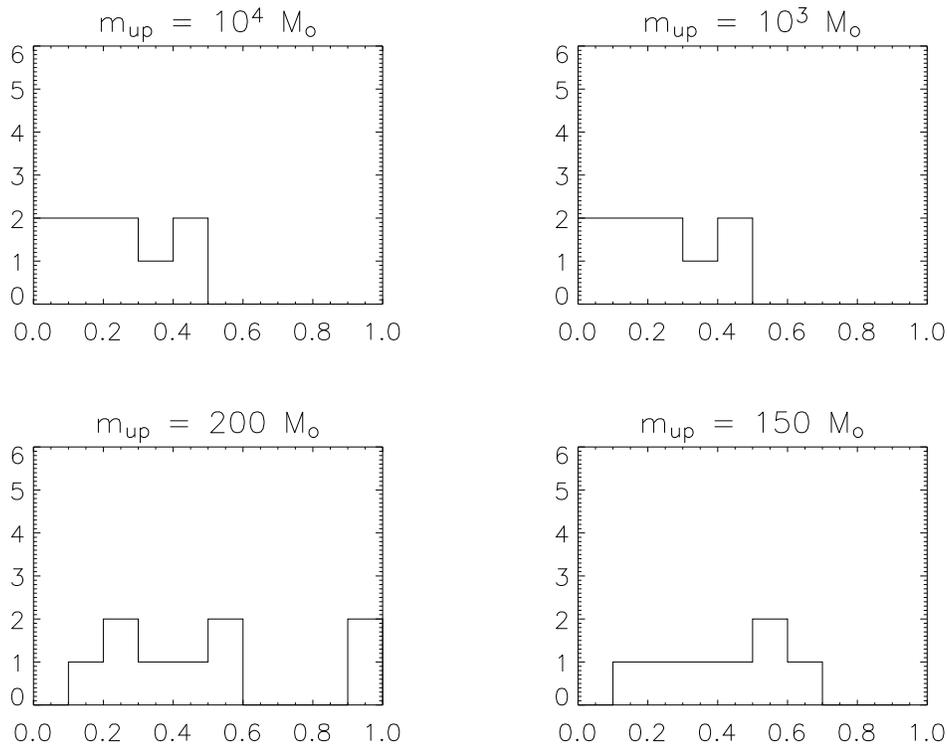}{3.8in}{0}{60}{60}{-230}{-160}
\caption{Histograms of $p(m_{\rm max})$ for assumed
  $m_{\rm up}=10^4$, 200, 150 and 120 M$_\odot$,
  as shown.  For unbiased conditions, $p(m_{\rm max})$ should show a
  uniform distribution.
\label{f_pmmax}
}
\end{figure}

However, given that the preceding is based on only 9 clusters, we may
worry that, because of stochasticity, these may not be representative
of the massive star population.  We can quantify the degree to which
we might be this unlucky by evaluating the probabilities $p(m_{\rm max})$ of
obtaining the observed $m_{\rm max}$ seen in each cluster, for an assumed $m_{\rm up}$.
Under normal circumstances, we should be uniformly lucky and unlucky,
and so these $p(m_{\rm max})$ should represent a uniform distribution when
$m_{\rm up}$ corresponds to its actual value in the parent distribution.
Figure~\ref{f_pmmax} shows the histograms of $p(m_{\rm max})$ for assumed
$m_{\rm up} = 10^4$, 200, 150, and 120 M$_\odot$.  The probabilities
$P$ that these histograms are drawn from uniform distributions for
these respective cases are $P < 0.002,\ < 0.02,\ 0.12,$ and $< 0.47$.
Thus we confirm that, only when $m_{\rm up}$ has values similar to the
observed $m_{\rm max}$, do we see significance in $P$, demonstrating that
$m_{\rm up}$ in this particular sample is indeed around 150 M$_\odot$ at the
significance implied by $P$.
We note that \citet{Aban2006} point out that the statistical
estimator for the maximum value of a truncated, inverse power-law
distribution is indeed the maximum value in the dataset, corresponding
to the observed $m_{\rm max}$.

It is especially remarkable that this apparent value of $m_{\rm up} \sim 150$
M$_\odot$ is seen over a great range in star-forming conditions.  It
holds for both ordinary OB associations in both the Milky Way, a massive,
spiral galaxy; and the LMC, a lower-mass, irregular galaxy with
presumably a lower-pressure ISM.  At the same time, the same
upper-mass limit applies in the R136a super star cluster, an extreme
environment in the LMC.  And \citet{Figer2005} finds the same
upper-mass limit in the Arches cluster near the Galactic Center, which
is another extreme, yet very different, environment.  We caution that
these results are based on the inferred stellar masses for these
systems, taken at face value, and assuming that the highest-mass stars
have not yet expired.

\citet{Koen2006} used the same dataset for R136a \citep{MasseyHunter1998} to
evaluate $m_{\rm up}$ and the IMF slope $\gamma$ from the cumulative stellar
mass function.  The cumulative distribution more clearly reveals the
existence of a truncated $m_{\rm up}$.  Koen simultaneously fit
$m_{\rm up}$ and $\gamma$ for both sets of masses provided by Massey \&
Hunter, based on two different calibrations for spectral type
to stellar effective temperature.  The fit was carried out with both a
least-squares method and a maximum likelihood method.  Across all four
cases, the results again are largely consistent with a Salpeter slope
and $m_{\rm up}\sim 150$ M$_\odot$.  Fixing $m_{\rm up}=\infty$ results in a very
steep slope, $\gamma\sim 4$ to 5, strongly inconsistent with the
Salpeter value.  An infinite $m_{\rm up}$ is eliminated with a probability
of $<0.0005$ and $<0.002$ for the least-squares and maximum-likelihood
methods, respectively, consistent with our results (Oey \& Clarke
2005).

\section{Sparse OB Star Groups}

Given the apparent robustness of the upper-mass limit in a variety of
local environments, we seek to examine other extreme
circumstances.  An especially interesting case is field
OB stars.  While a substantial fraction of these are likely to be
runaway stars ejected from clusters, many may well be members of
small, low-mass clusters.  \citet{Oey2004} demonstrated that
individual SMC field OB stars, as defined by a friends-of-friends
algorithm, do fall smoothly on a continuous power-law distribution in
the number of stars per cluster $N_*$, which is akin to the cluster mass function.
This supports the scenario that most field OB stars are members of
low-mass clusters.

We obtained F555W and F814W SNAP observations of eight, apparently
isolated OB stars from among these SMC field stars, using the {\it HST}
ACS camera \citep{Lamb2010}.  Applying both a stellar density
analysis and a friends-of-friends algorithm, we confirmed that three
of these objects have stellar density enhancements, corresponding to
sparse stellar groups.  An additional object registered a positive
signal for companions using the friends-of-friends algorithm, but not
the stellar density analysis.  The remaining four stars appear
isolated to within the detection limit of F814W = 22 mag.  Radial
velocity observations subsequently revealed two of these 
stars to be runaway stars.  This leaves two stars remaining
as candidates for {\it in-situ} isolated, field OB stars.

We constructed Monte-Carlo simulations of cluster populations to
examine the parameter space occupied by our observed, sparse OB star
groups.  We first adopted a cluster mass function (MF) given by a simple
power-law distribution:
\begin{equation}\label{cmf}
N(M_{\rm cl})\ dM_{\rm cl} \propto M_{\rm cl}^{-2}\ dM_{\rm cl} \quad ,
\end{equation}
where $N(M_{\rm cl})$ is the number of clusters in the mass range
$M_{\rm cl}$ to $M_{\rm cl} + dM_{\rm cl}$.  Each cluster drawn
from this distribution is then populated with a stellar IMF drawn from
a \citet{Kroupa2001} IMF, which has the form,
\begin{equation}
n(m)\ dm \left\{
\begin{array}{ll}
m^{-1.3}\ dm\ , \quad 0.08 {\rm M_\odot}\leq m < 0.5\ {\rm M_\odot} \\
m^{-2.35}\ dm\ , \quad 0.5 {\rm M_\odot}\leq m < 150\ {\rm M_\odot} 
\end{array}
\right.
\end{equation}
We adopt a default model having a power-law index of --2
for the cluster MF (equation~\ref{cmf}), and lower-mass limit of $M_{\rm cl,lo} =
20$ M$_\odot$.  This model best reproduces the frequency of single-O star
clusters in the SMC and Milky Way (see Lamb et al. 2010).  

\begin{figure}
\plotone{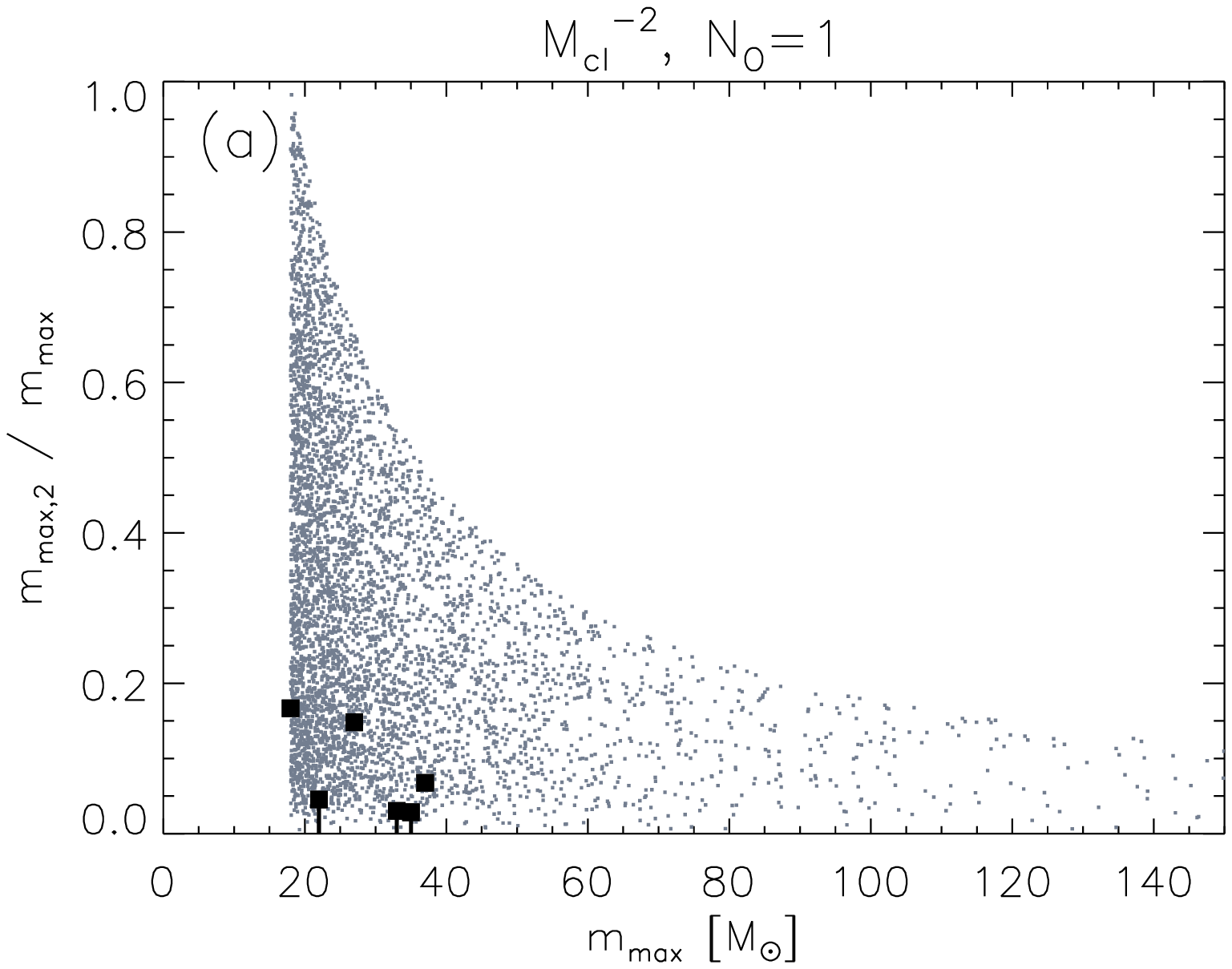}
\plotone{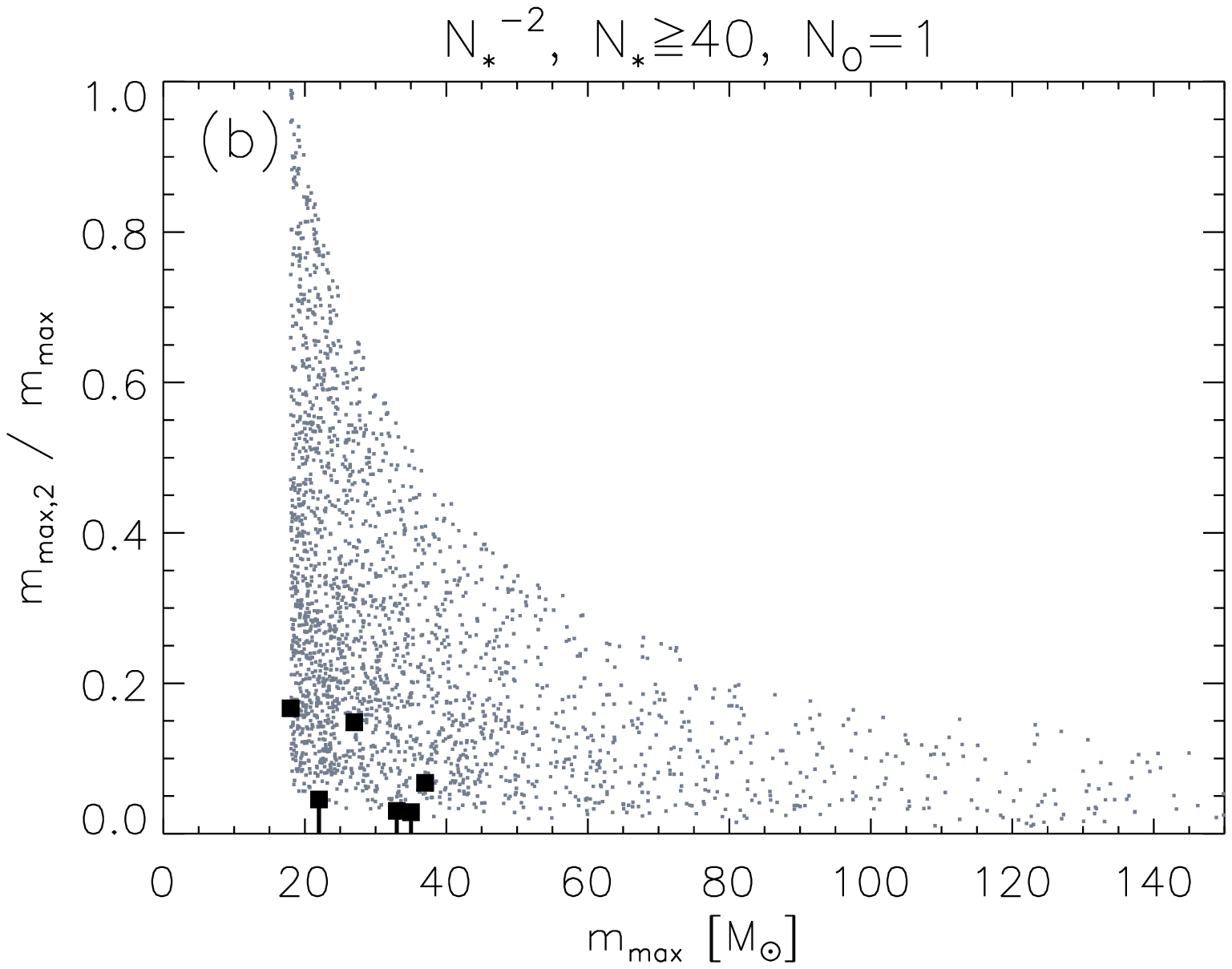}
\caption{Mass ratio of the two highest-mass stars vs $m_{\rm max}$ for
  cluster simulations based on a cluster MF (panel $a$; top) and on a
  cluster membership function (panel $b$; bottom).  Only simulated clusters
  having a single OB star are shown, with gray points.  Black squares correspond to our
  observed sparse O star groups, with three objects showing upper limits.
\label{mrat}
}
\end{figure}

Figure~\ref{mrat}$a$ shows the mass ratio distribution of the
second-highest to highest mass stars $m_{\rm max,2}/m_{\rm max}$ vs
$m_{\rm max}$ in these simulations.
Only clusters having a single OB star, defined in this
figure as having mass $m\ge 18$ M$_\odot$ are shown.  Our observed objects are
overplotted with the black squares, showing upper limits for the
apparently isolated objects.
We see that while our observed objects appear to fall in a
densely-populated region of the parameter space, the simulations peter
out at the lowest values of $m_{\rm max,2}/m_{\rm max}$,  
since the probability of drawing an extremely low mass ratio is
tiny.  All of our data fall within the lowest 20th percentile in 
$m_{\rm max,2}/m_{\rm max}$, and these frequencies also result when our clusters are
simulated based on cluster membership number $N_*$ instead of a MF
(Figure~\ref{mrat}$b$).  This model analogously uses the form
$N_*^{-2}$, the Kroupa IMF, and a lower limit of $N_{\rm *,lo} = 40$
stars, which is equivalent to $M_{\rm cl,lo}$ used above.
That our objects all fall in this low-frequency regime is partly due
to our selection of apparently isolated OB stars as the {\it
  HST} targets.  We note that the simulations reproduce
this regime fairly easily.

\begin{figure}
\plotone{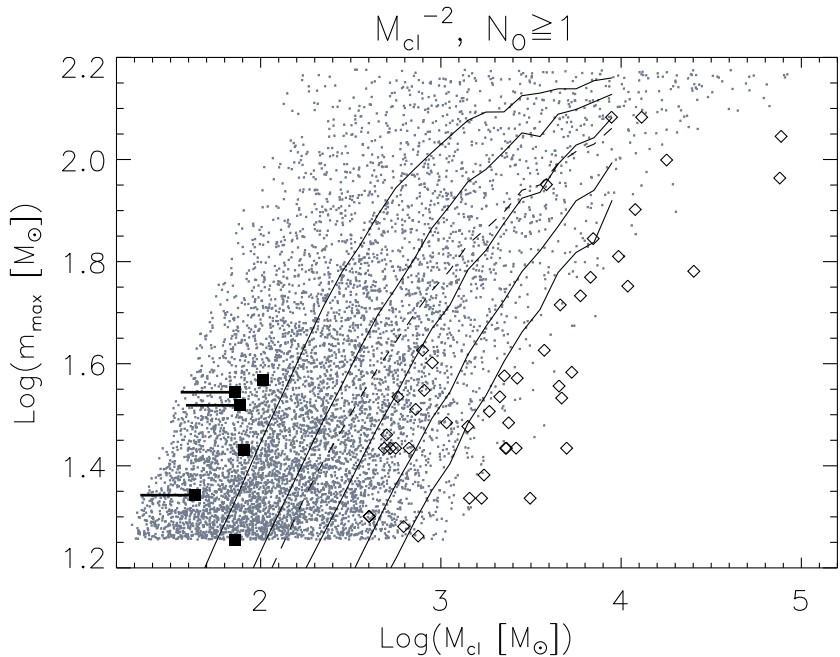}
\caption{$m_{\rm max}$ vs $M_{\rm cl}$ in
  simulations based on a cluster MF, for all objects with one or more OB stars.
  Our observed objects and the sample compiled by
  \citet{Weidner2010} are shown in black squares and open diamonds,
  respectively.  Solid lines show the 10th, 25th, 50th, 75th, and 90th percentiles of
  $m_{\rm max}$, and the dashed line shows the mean.
\label{mmaxmcl}
}
\end{figure}

Figure~\ref{mmaxmcl} shows the relation between the $m_{\rm max}$ and $M_{\rm cl}$ for
the simulations based on the cluster mass function
(equation~\ref{cmf}), now showing all clusters having at least one OB
star.  Since there is no relation imposed between the $m_{\rm max}$ and $M_{\rm cl}$,
the simulations occupy a large parameter space.  The solid lines
show the 10th, 25th, 50th, 75th, and 90th percentiles of $m_{\rm max}$ for a
given $M_{\rm cl}$, and the dashed line shows the mean, equivalent to
equation~\ref{eqmmax}.  Our observed objects 
are shown with the black squares as before, with $M_{\rm cl}$ computed by
assuming that a fully populated IMF exists below the detection
threshold of $\sim 1.5\ {\rm M_\odot}$.  Open diamonds show the
sample of observed clusters compiled by \citet{Weidner2010}.

We see that for almost all of our observed objects, 90\% of the simulated
clusters have lower $m_{\rm max}$ for a given $M_{\rm cl}$, whereas
the sample from Weidner et al. (2010) is largely 
at the opposite extreme.  Our data, taken at face value, are in
substantial conflict with the premise of a well-defined $m_{\rm max}$ -- $M_{\rm cl}$
relation, proposed by \citet{WeidnerKroupa2005}.  The observed data
show a scatter spanning two orders of magnitude in $M_{\rm cl}$ for the $m_{\rm max}$
in our sparse OB groups.  While it is essential to understand the
consequences of cluster dynamical evolution, this range in values,
which is also consistent with the unconstrained simulations, 
strongly suggests that a $m_{\rm max}$ -- $M_{\rm cl}$ relation is
weak at best, as also found by \citet{MaschbergerClarke2008}.
\citet{Testi1999} discovered sparse groups around Galactic Herbig
Ae/Be stars, which are newly formed objects, and thus their parent
systems are unlikely to be strongly depleted by dynamical evaporation.
Our observations extend such findings to OB stars.

The existence of a relation between $m_{\rm max}$ and $M_{\rm cl}$ is widely debated.
A clear relation would strongly impact the integrated galaxy IMF
(IGIMF; Weidner \& Kroupa 2005), affecting interpretations of stellar populations,
star-formation histories, and inferred galaxy evolution.  Moreover,
competitive accretion theories for star formation predict the
existence of such a relation:   $m_{\rm max}\propto M_{\rm cl}^{2/3}$
\citep{Bonnell2004}.  It is therefore of great interest to further
investigate whether sparse OB star groups are representative of their
birth conditions.

\section{Summary}

With samples of uniformly derived OB star masses, we can begin to
quantitatively evaluate the properties of the stellar upper IMF, based on
a stochastic understanding of its nature.  We demonstrated the
existence of a stellar upper-mass limit $m_{\rm up}\sim 150$ M$_\odot$ for a
Salpeter slope, using data for the massive star census in a sample
of ordinary Milky Way and LMC OB associations, and the super star
cluster R136a in 30 Doradus (Oey \& Clarke 2005).  
Based the probability of observing the highest-mass stars in their
respective clusters, we confirmed the existence of a strong deficit
above this value of $m_{\rm up}$.  Koen (2006) examined the
cumulative distribution function of the stellar masses, and by
jointly fitting $m_{\rm up}$ together with a power-law
distribution for R136a, he confirmed a value for $m_{\rm up} \sim 150$ M$_\odot$
and a Salpeter-like slope as the best-fit values for these parameter
estimations.  The variety of environments in which this value of the
upper-mass limit applies is remarkable:  ordinary OB associations, the
R136a super star cluster, and the Galactic Center environment.

To further examine the robustness of this upper-mass limit, we
searched for low-mass companions around eight SMC field OB stars with
{\it HST}/ACS.  We confirmed the existence of sparse groups
associated with 3 -- 4 of these field massive stars, while 2 targets are
runaway OB stars, and 2 -- 3 remain candidates for isolated OB stars
that formed {\it in situ} (Lamb et al. 2010).  We generate Monte Carlo
simulations of cluster populations based on simple power-law sampling
for both cluster and stellar masses, and we find that the cluster 
lower-mass limit is most consistent with the frequency of SMC and
Galactic field O stars at $M_{\rm cl,lo} \sim 20$ M$_\odot$ or 
$N_{\rm *,lo}\sim 40$ for a Kroupa IMF.  Our sparse
OB groups all fall in the lowest 20th percentile
in $m_{\rm max,2}$/$m_{\rm max}$, regardless
of whether the clusters are populated by $M_{\rm cl}$ or by $N_*$.    

Our sparse OB groups generally fall in the highest 10th percentile of
$m_{\rm max}$ for a given $M_{\rm cl}$, at an opposite extreme from the sample of
objects compiled by Weidner et al. (2010), which fall in the locus
of the most massive objects, at a given $m_{\rm max}$, in our simulations.
Taken at face value, our results therefore contradict the existence of
a well-defined relation between $m_{\rm max}$ and $M_{\rm cl}$.  If our sample is not
predominantly the result of dynamical evaporation, then this finding
may pose difficulties for the predicted steepening of the IGIMF based
on a suggested $m_{\rm max}$ -- $M_{\rm cl}$ relation (e.g., Weidner \& Kroupa 2005)
and competitive accretion theories for star formation (Bonnell et
al. 2004), which also rely on the existence of such a relation.

\acknowledgements 

This work was supported by NASA HST-GO-10629.01 and NSF grant AST-0907758.

\bibliography{Oey_MS}

\end{document}